\documentstyle[11pt]{article}
\input epsf
\begin{document}

\baselineskip 22pt
\thispagestyle{empty}
\rightline{CU-TP-683}
\vskip 1cm
\centerline{\Large\bf Vanishing Hawking Radiation from}
\centerline{\Large\bf a Uniformly Accelerated Black Hole}
\vskip 1.5cm
\centerline{Piljin Yi \footnote{e-mail address:
piljin@cuphyc.phys.columbia.edu}}
\vskip 5mm
\centerline{Physics Department, Columbia University}
\centerline{New York, New York, 10027, U.S.A}
\vskip 1.5cm
\centerline{\bf ABSTRACT}
\vskip 5mm
\begin{quote}
We consider quantum fields around uniformly accelerated black holes in the
eternal Ernst geometry. At a particular
value of the acceleration, the Bogolubov
transformation which would be responsible for the late-time Hawking radiation,
is found to be trivial. When this happens, Hawking's thermal radiation,
Doppler-shifted or not, is absent to the asymptotic inertial observers despite
the nonzero Hawking temperature, while the co-moving observers find the black
hole radiance exactly balanced by the acceleration heat bath. We close with
a few comments.
\end{quote}
\vskip 1.5cm
\centerline{\it Physical Review Letters \ \ vol. 75 (1995) 382}

\vskip 2cm
\leftline{PACS\#: 04.70.Dy, 03.70.+k, 04.20.Gz}
\newpage

The Hawking radiation from the black hole at rest \cite{hawking}, is
by now a well-understood phenomenon within the semiclassical framework. Many
quantum mechanical concepts, such as energy quanta and the occupation numbers,
turned out to be coordinate-dependent ones, and this ambiguity leads to the
particle creation in the presence of the event horizon. A black hole
that has nonzero $T_{BH}$, emits thermal radiations and thereby loses its
mass steadily, unless it is stabilized by a conserved local charge inside.
The canonical example of the latter is given by the well-known
Reissner-Nordstrom (RN) black holes \cite{gravity}, the minimal variant of
which, namely the extremal case, has zero Hawking temperature.

A related phenomenon of some interest is the so-called acceleration heat
bath \cite{bath}\cite{unruh}. Through a similar quantum field theoretical
effect as above, the usual Minkowskian vacuum feels like a heat bath to
uniformly accelerated observers (sometimes called Rindler observers), with the
acceleration temperature $T_{A}$ being equal to the acceleration multiplied
by $\hbar/2\pi$.

Now a puzzling question that follows \cite{piljin}, is what happens to
a uniformly accelerated black hole when the nonzero Hawking temperature
$T_{BH}$ equals $T_{A}$. One naive expectation would be that inertial observers
will find Doppler-shifted Hawking radiation (since the acceleration heat bath
is not real to these observers), and that the black hole continues to
evaporate. However, if one follows the black hole at large fixed distance so
that he himself undergoes the same acceleration, one must find that the
thermal radiation from the black hole is exactly balanced by the acceleration
heat bath, and that the black hole does not evaporate at all.

One might be tempted to dismiss this puzzle \cite{giddings} as a variant of
the famous problem of a uniformly accelerated charge. In the latter system,
co-moving Rindler observers again find a time-independent equilibrium state,
yet inertial observers detect net electromagnetic radiation energy
\cite{boulware}. Is there a way to reconcile two such potentially conflicting
observations in our case also?

Later, we will find that, unlike the case of a uniformly accelerated charge,
both groups of observers actually agree in our case: No net Hawking radiation
emanates from the black hole. The respective physical explanations are
different, however. The Rindler observers find a co-moving heat bath that
continually exchanges quanta with the black hole, in accordance with the
previous expectation. On the other hand, no particle-creation (analogous to
the usual Hawking radiation) occurs at all to the asymptotic inertial
observers, contrary to the naive expectation.

In order to facilitate the derivation, we will specialize to the static Ernst
geometry \cite{ernst}, where a pair of near-extremal magnetic RN black
holes are undergoing an eternal uniform acceleration. Although this enabled us
to derive the desired result most clearly  and succinctly, the conclusion
appears quite independent of such details as whether the acceleration is
eternal, in much the same way as one can derive the Bogolubov transformation
relevant for Hawking's radiation in the geometry of eternal black holes.
See Ref.~\cite{next} for more detail, as well as for a toy model where explicit
estimates are made with more realistic initial condition.

\vskip 1cm
\begin{center}
\leavevmode
\epsfysize 3in \epsfbox{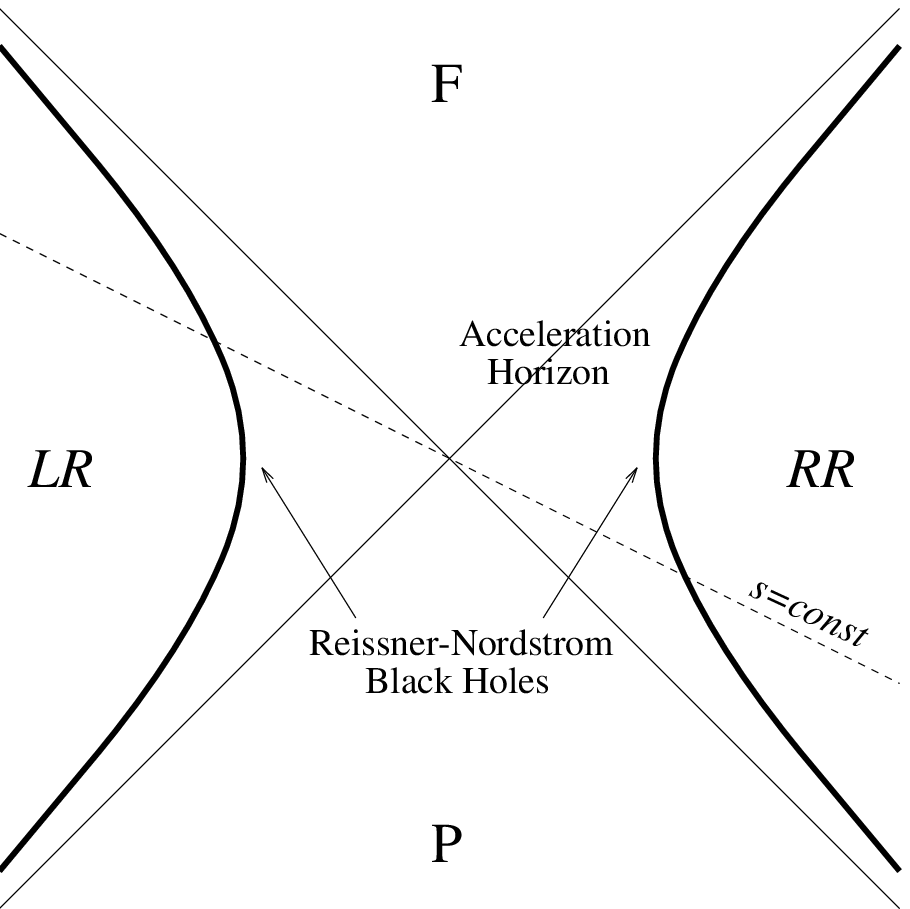}
\end{center}
\begin{quote}
{\bf Figure 1:} {\small A schematic diagram for a pair of uniformly
accelerated black holes. The black holes are represented by two
hyperbolic world lines in each Rindler wedges.}
\end{quote}

\vskip 5mm
Let us first write down the Ernst metric in a new coordinate system.
\begin{equation}
g=\frac{\Lambda^2}{(1+rAx)^2}\biggl\{-F(r)\,ds^2+F(r)^{-1}dr^2+r^2\,G(x)^{-1}
dx^2+r^2\,G(x)\,\Lambda^{-4}d\phi^2\biggr\}. \label{metric}
\end{equation}
The Killing time coordinate $s$ is actually a Rindler-type
coordinate, so that the ``static'' observers are uniformly accelerated and
thus, along with the black holes, must be confined within the Rindler
wedges {\large\it LR} and {\large\it RR} in figure 1. The spatial
coordinate $r$ plays the role of the usual radial coordinate
only near the black hole horizon, as is easily seen from the form of $F(r)$.
\begin{equation}
F(r)\equiv -A^2r^2\,G(-1/Ar)=(1-\frac{r_-}{r})(1-\frac{r_+}{r}-A^2r^2).
\end{equation}
This also shows that the geometry resembles that of the static RN black holes
wherever $Ar$ is sufficiently small. The external magnetic field that drives
the acceleration is encoded in $\Lambda$:
\begin{equation}
\Lambda\equiv \biggl\{1+\frac{Bx}{2}\sqrt{r_+ r_-}\,\bigg\}^2+\frac{
B^2r^2}{4 (1+rAx)^2}\,G(x),
\end{equation}
where $B$ is approximately the magnetic field strength that drives the uniform
acceleration.

Note that the same quartic polynomial $G$ appears in all components of the
metric. Call the four roots of it, $\xi_1$, $\xi_2$, $\xi_3$, $\xi_4$ in the
ascending order. Then, the event horizon of the black hole is at
$r=\tilde{r}_+ \equiv-1/\xi_2A$, while $r=r_{A}\equiv -1/\xi_3A$ is the
acceleration horizon. Define the surface gravities of the horizons:
\begin{equation}
\kappa_{BH}\equiv \frac{F'(\tilde{r}_+)}{2},\qquad
\kappa_{A}\equiv -\frac{F'(r_{A})}{2},
\end{equation}
which are related to the temperatures by $T_{BH}=\hbar\kappa_{BH}/2\pi$ and
by $T_A=\hbar\kappa_{A}/2\pi$. When the size of the black holes are
relatively small ($r_\pm A \ll 1$), $A\simeq \kappa_A$ can be regarded as
the acceleration of the black hole.

Since we are interested in the cases where the co-moving Rindler observers
find a complete thermal equilibrium, we want to require that the Hawking
temperature be equal to the acceleration temperature. In terms of the surface
gravities, therefore, we demand that
\begin{equation}
\kappa\equiv \kappa_{BH}=\kappa_{A}.\label{kappa}
\end{equation}
In some cases, most notably when the black hole mass is
much larger that its charge, $\kappa_{BH} >\kappa_A$ is always true and this
constraint can never be met. However, when the non-extremal RN black holes in
question are sufficiently close to the extremality, it is possible to achieve
this fine-tuning \cite{piljin}. In fact, this constraint is naturally imposed
if the two black holes are pair-created via the wormhole-type instanton
\cite{garfinkle}.

\vskip 5mm
In addition to the above considerations, it is most essential for our purpose
that we understand the causal structure of the space-time. First of all,
recall that the acceleration horizon divides the asymptotic infinities into
two groups: ones inside Rindler wedges and ones outside. Inside the Rindler
wedges {\it LR} and {\it RR}, there are asymptotic null or
space-like infinities that correspond to $x=\xi_3=-1/Ar$. However, most
spacetime trajectories of energy quanta, being either time-like or null, will
eventually cross acceleration horizon into the region {\large F} which is
inaccessible to those ``static'' Rindler
observers \cite{boulware}. In order to
remain within the Rindler wedge forever, the particle must either have
magnetic charge or be directed exactly parallel to the axis of the uniform
acceleration. It follows that {\it in order to understand
the radiative process, one may safely ignore the
infinities in the Rindler wedges.} The resulting causal structure can be
found in figure 2 below. Throughout this letter, we shall assume that there
exist no ``light'' magnetically charged particles.
\vskip 1.5cm

\begin{center}
\leavevmode
\epsfysize 4in \epsfbox{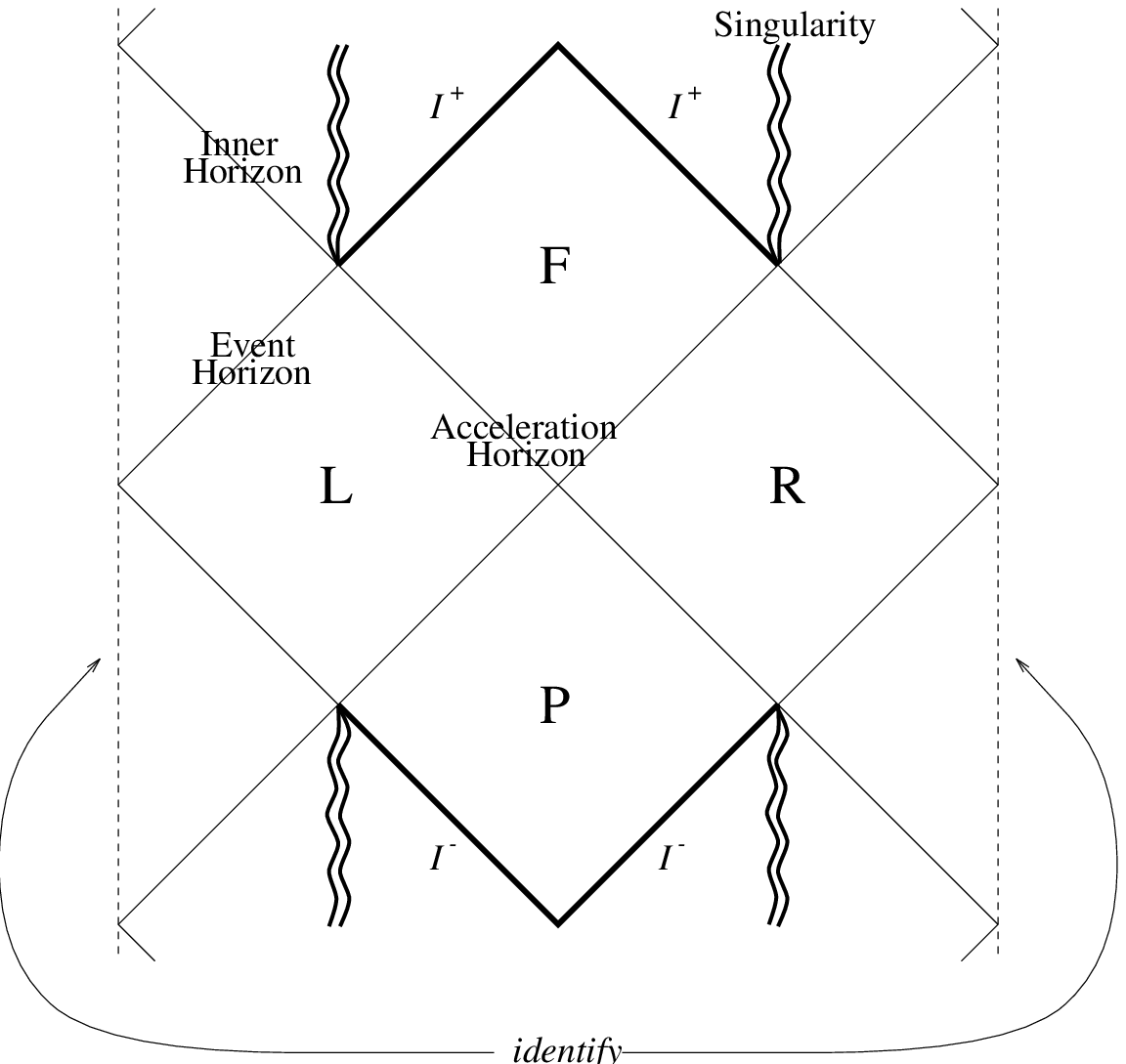}
\end{center}
\begin{quote}
{\bf Figure 2:} {\small Penrose diagram of the Ernst spacetime
with the Rindler infinities at $x=\xi_3=-1/Ar$ excised. The bold (straight)
lines indicate the asymptotic infinities.}
\end{quote}

\vskip 5mm

This spacetime actually represents a wormhole. In figure 2, the
left end of the diagram is identified with the right end. However, this
wormhole is different from those found in eternal black holes \cite{gravity}
in one important aspect: Here, the two mouths (or equivalently the two
black holes) share the same asymptotic future {\large F}.

\vskip 5mm
As a first step in quantizing matter fields, consider the field equation for
a free scalar in regions {\large  L} and {\large R}. For this purpose,
it is most convenient to introduce a new tortoise-like coordinate $z$ between
the two horizons ($\tilde{r}_+\le r \le r_A$):
\begin{equation}
z\equiv \int^r d\tilde{r}\,\frac{1}{F(\tilde{r})},
\end{equation}
which logarithmically approaches $-\infty$ at the event horizon and $+\infty$
at the acceleration horizon. After rescaling the eigenmodes $\Psi^{(w,m)}$,
for each Rindler frequency $w>0$ and the quantized angular momentum $m$,
\begin{equation}
\Psi^{(w,m)}= e^{\mp iws}\,\frac{(1+rAx)}{r}\,[\,\Phi_{(w,m)}(r,x)
e^{im\phi}\,],
\end{equation}
we find the following equation that must be solved for the eigenmodes:
\begin{equation}
w^2\Phi_{(w,m)}+\frac{\partial^2}{\partial z^2}\Phi_{(w,m)}=F(r(z))\biggl\{
\frac{1}{r^2}\biggl[-\frac{\partial}{\partial x} G(x)\frac{\partial}{\partial
x} +\frac{m^2\Lambda^4}{G(x)}\biggr]+U_{\rm eff} \biggr\}\,\Phi_{(w,m)}.
\label{eigen}
\end{equation}
$U_{\rm eff}$ is a bounded function of $z$ and $x$, and in
particular contains the mass term and possible curvature couplings. Note that
the right-hand-side of Eq.~(\ref{eigen}) has the overall factor $F(r(z))$
that vanishes exponentially $\sim e^{-2\kappa |z|}$ as $|z|\rightarrow
\infty$. In the same limit, $\Lambda$ is reduced to functions of $x$ only,
$\Lambda_A \equiv \Lambda(r=r_A)$ or $\Lambda_{BH}\equiv\Lambda(r=
\tilde{r}_+)$.

Introducing two null coordinates $u=s-z$ and $v=s+z$, we find the following
general behavior, near each horizon, of the future-directed Rindler eigenmodes
$\Psi^{(w,m)}_L$ and $\Psi^{(w,m)}_R$ that have respective supports in either
{\large L} or {\large R},
\begin{eqnarray}
\Psi_L^{(w,m)} \sim e^{-iwu}C_{\lambda m}(x) e^{im\phi}\hbox{ \ or }
e^{-iwv}C_{\lambda m}(x) e^{im\phi}\quad \hbox{ in L},& & \qquad
\Psi_L^{(w,m)} =0 \quad \hbox{ in R }, \label{modeL}\\
\Psi_R^{(w,m)} \sim e^{+iwu}C_{\lambda m}(x) e^{im\phi}\hbox{ \ or }
e^{+iwv}C_{\lambda m}(x) e^{im\phi}\quad \hbox{ in R},& & \qquad
\Psi_R^{(w,m)} =0 \quad \hbox{ in L }. \label{modeR}
\end{eqnarray}
The positive sign in (\ref{modeR}) is because $(u,v)$ grow toward past
rather than future in the region {\large R}.
The same set of symbols $C_{\lambda m}$ and $\lambda$ are used to denote
eigenfunctions and eigevalues for two different eigenvalue problems, the
relevant operators of which are obtained from the one inside the
square bracket in Eq.~(\ref{eigen}), by replacing $\Lambda$ by $\Lambda_A$ or
by $\Lambda_{BH}$. In particular, due to the lack of the spherically
symmetry, an eigenmode that has a definite $\lambda$ near the event
horizon will not have a definite $\lambda'$ near the acceleration horizon.
Similarly, the backscattering effect will mix the left-moving modes with the
right-moving modes. But since none of these details matter, as we will find
out shortly, we shall keep just one superscript $w$ form now on.

\vskip 5mm
Let us be reminded that, for each positive Rindler mode $\Psi^{(w)}$
with $w >0$, there exists a time-reversed negative mode $\Psi^{(-w)}$
that propagates backward but otherwise of the same form: The complete Hilbert
space is spanned by both positive and negative modes. But the point is, such
labels as future-directed and past-directed are inherently observer-dependent.
A purely future-directed (positive) mode in one coordinate system could be
a mixture of both future-directed (positive) and past-directed (negative)
modes as perceived in another. Accordingly, the so-called {\it Bogolubov
transformation}, which maps one basis to the other, is such that a vacuum with
respect to one set of observers can actually be an excited state with respect
to the other. And this is exactly the origin of both the Hawking radiation
and the acceleration heat baths \cite{hawking}\cite{unruh}.

In this regard, it is important to realize that $(u,v)$ are not good
coordinates for inertial observers and must be traded off in favor of
Kruskal-type coordinates that play the role of advanced and retarded times
for local inertial observers. The approximate form of such coordinates
near the respective horizons, are completely determined by the surface gravity
$\kappa$ alone: Calling the Kruskal coordinates near the event horizon
$(U_1,V_1)$, we find,
\begin{eqnarray}
\kappa u\simeq - \ln (-U_1) \quad\hbox{ in L},& &\qquad
\kappa u\simeq - \ln (+U_1) \quad\hbox{ in R}, \label{uU}\\
\kappa v\simeq + \ln (+V_1) \quad\hbox{ in L},& &\qquad
\kappa v\simeq + \ln (-V_1) \quad\hbox{ in R}. \label{vV}
\end{eqnarray}
For the other Kruskal coordinates $(U_2,V_2)$ near the acceleration horizon,
we simply replace $(U_1,V_1)$ by $(U_2,V_2)$ and reverse every single sign on
the right-hand-side.

At last, we are ready to obtain the eigenmodes of positive frequencies
with respect to inertial observers. For the purpose, we may use Unruh's
characterization of positive frequency \cite{unruh}: a simple
analyticity argument shows that a positive frequency mode must be
analytic and bounded in the lower-half-plane of the complexified time
coordinate. For instance, a positive frequency mode as detected by the
inertial observers near the event horizon, must be analytic in $U_1$
and $V_1$ throughout their lower-half-planes.

Since the Rindler modes are defined in either {\large L} or {\large R}, they
are defined only on the half-lines of Kruskal coordinates. To construct the
eigenmodes that are appropriate for inertial observers, one expresses
$\Psi^{(w)}_L$ and $\Psi^{(w)}_R$ in terms of $(U_i,V_i)$ for $i=1,2$ using
the coordinate transformations above in (\ref{uU}) and (\ref{vV}), and
analytically continue the logarithms through lower-half-planes of each
Kruskal coordinates. Then the resulting modes have positive frequencies with
respect to inertial observers, in addition to having the supports on the
entire spans of Kruskal coordinates. For inertial observers
near the event horizon of the black holes, the positive frequency modes
${\Psi}_B^{(w)}$ are
\begin{equation}
{\Psi}^{(w)}_{BL}\simeq N_w[\Psi^{(w)}_L+e^{-\pi w/\kappa} \Psi^{(-w)}_R],
\qquad
{\Psi}^{(w)}_{BR}\simeq N_w[\Psi^{(w)}_R+e^{-\pi w/\kappa} \Psi^{(-w)}_L],
\label{RB}
\end{equation}
where $N_w\equiv 1/\sqrt{1-e^{-2\pi w/\kappa}}$. The expressions for
$\Psi^{(-w)}_B$ can be obtained similarly.

For an ordinary nonaccelerated black holes ($\kappa=\kappa_{BH}$,
$\kappa_A=0$), this would be the end of the story, since the Rindler
coordinates $(u,v)$ are in that case the ordinary retarded and advanced
times asymptotically: the positive Rindler modes $\Psi^{(w)}$ themselves are
future-directed with respect to asymptotic inertial observers.
For physical black holes with smooth future event horizon, then, the
physical vacua are such that the asymptotic inertial observers will
find outward thermal radiation at $T_{BH}=\hbar\kappa_{BH}/2\pi$
\cite{hawking}.

\vskip 5mm
However, with the uniformly accelerated black hole, $(u,v)$ are not good
asymptotic inertial coordinates. Rather, $(U_2,V_2)$ are. Far away from
the black hole, the metric (\ref{metric}) is essentially that of the Melvin
spacetime in a Rindler-type coordinate system \cite{giddings},
which means that $(V_2+U_2)/2$ may be regarded as the asymptotic Minkowski
time. This is particularly clear when the black holes are relatively small
($r_\pm A \ll 1$) \cite{piljin}.

We must perform another Bogolubov transformation and superimpose it to the
one in Eq.~(\ref{RB}). Near the acceleration horizon, the situation is
identical to the above (because $\kappa_A=\kappa\equiv\kappa_{BH}$), except
that the relative positions of {\large L} and  {\large R} are switched.
Calling the asymptotic inertial modes $\Psi_A^{(w)} $'s, we find near the
acceleration horizon:
\begin{equation}
{\Psi}^{(w)}_{AR}\simeq N_w[\Psi^{(w)}_R+e^{-\pi w/\kappa} \Psi^{(-w)}_L],
\qquad
{\Psi}^{(w)}_{AL}\simeq N_w[\Psi^{(w)}_L+e^{-\pi w/\kappa} \Psi^{(-w)}_R].
\label{RA}
\end{equation}
Combining the two transformations in (\ref{RB}) and (\ref{RA}), we finally
find that {\it the Bogolubov transformation} between the two groups of
inertial observers, {\it is actually  trivial.} That is, it does not mix
positive and negative frequency modes:
\begin{equation}
\Psi^{(w)}_B\Rightarrow \Psi^{(w)}_A,\qquad
\Psi^{(-w)}_B\Rightarrow \Psi^{(-w)}_A. \label{main}
\end{equation}
As was noted earlier, the respective forms of the Rindler modes near each
horizon must be taken with a grain of salt: Combining the two Bogolubov
transformations, in general, we must include a unitary transformation $\cal
U$ that reflects the
effects of local geometry between the two horizons. On the other hand, the
Rindler time $s$ is a Killing coordinate, and $\cal U$ has to commute with
the Rindler energy operator $i\hbar\,\partial_s$. Hence, with the Bogolubov
transformations above that depend only on the Rindler frequency $w$, this
additional complication cannot alter our conclusion: {\it we find that no
late-time Hawking radiation reaches the asymptotic inertial observers}.
\vskip 5mm

This may be compared to the classic puzzle of the Bremmstrahlung from a
uniformly accelerated charge. There too, the energy of the
moving charge is not diminished in any way, thanks to the vanishing
radiation backreaction, yet asymptotic inertial observers find {\it
classical} electromagnetic radiation emanating from the charge. This
energetics part of the puzzle was first resolved by Coleman \cite{coleman}
in part and then later by Boulware \cite{boulware} more completely,
who observed that the radiation actually originates from the conversion
of the ``pre-acceleration'' Coulomb field to the one associated with the
moving charge. If the uniform acceleration is terminated in the future, the
charge will give up some of its kinetic energy to restore the Coulomb field
again, but as long as the uniform acceleration is maintained, all the radiation
energy is derived from the continuously varying electromagnetic field around
the charge. The obvious difficulty anyone must face in trying to find a
similar mechanism in our problem, is simply that there is no such
``pre-acceleration'' field for each and every particle species: the Hawking
radiation is completely universal in its composition.

\vskip 5mm
We would like to close with a few comments. First, we wish to emphasize that
the so-called late-time approximation is employed here, as in most derivations
of black hole radiance. While the main result (\ref{main})
precludes any analogue of the late-time Hawking radiation, we still
may not be able to account for possible (sub-leading) transient behaviors.
In fact, such an effect that corrects the black hole mass by a finite
and small factor $\sim \hbar/r_+^2$, was previously observed for {\it
static} extremal RN black hole \cite{jaemo} directly as well as for
pair-produced near-extremal RN black holes \cite{piljin} indirectly.
However, once a steady state (with $\kappa_A=\kappa_{BH}$) is reached,
the main result (\ref{main}) tells us that no further radiation escapes into
the asymptotic future.

Second, we must point out that, although our method can be used to derive
the nontrivial Bogolubov transformation when $\kappa_A\neq \kappa_{BH}$, it
does not automatically provide the spectrum that the asymptotic
inertial observers detects. One of the reason is, the eigenmodes
$\Psi^{(w)}_A$ have a definite absolute value of energy (although not
the sign thereof) with respect to the Rindler time $s$, not with respect
to the the asymptotic inertial time $(V_2+U_2)/2$. (In our derivations above,
however, this fact was actually to our advantage, for it enabled us to
circumvent nasty complications that may arise from the Doppler effect.)
More direct approaches may be desirable.

\vskip 5mm

The author is grateful to Jaemo Park, J. Preskill and S. Trivedi for
interesting conversations at an early stage of this work. He also thanks
Kimyeong Lee, S.A. Ridgeway and E. Weinberg for reading a preliminary
version of the manuscript. This work is supported in part by U.S. Department
of Energy.

\end{document}